\documentclass[]{interact}

\usepackage{epstopdf}
\usepackage[caption=false]{subfig}
\usepackage{caption}
\usepackage{lipsum}
\usepackage{graphicx}
\usepackage[table,xcdraw]{xcolor}
\usepackage{subfloat}

\usepackage[toc,page]{appendix}
\usepackage{hyperref}

\usepackage[numbers,sort&compress]{natbib}
\bibpunct[, ]{[}{]}{,}{n}{,}{,}
\renewcommand\bibfont{\fontsize{10}{12}\selectfont}
\makeatletter
\def\NAT@def@citea{\def\@citea{\NAT@separator}}
\makeatother

\theoremstyle{plain}

\theoremstyle{definition}

\theoremstyle{remark}

\begin{document}

\title{A first look at COVID-19 information and misinformation sharing on Twitter}

\author{
\name{Lisa Singh\textsuperscript{a}\thanks{lisa.singh@georgetown.edu}, Shweta Bansal\textsuperscript{a}, Leticia Bode\textsuperscript{a}, Ceren Budak\textsuperscript{b},  Guangqing Chi\textsuperscript{c}, Kornraphop Kawintiranon\textsuperscript{a}, Colton Padden\textsuperscript{a}, Rebecca Vanarsdall\textsuperscript{a}, Emily Vraga\textsuperscript{d}, Yanchen Wang\textsuperscript{a} }
\affil{\textsuperscript{a}Georgetown University; \textsuperscript{b}University of Michigan;\\ \textsuperscript{c}Pennsylvania State University; \textsuperscript{d}University of Minnesota}
}

\maketitle

\begin{abstract}
Since December 2019, COVID-19 has been spreading rapidly across the world. Not surprisingly, conversation about COVID-19 is also increasing. This article is a first look at the amount of conversation taking place on social media, specifically Twitter, with respect to COVID-19, the themes of discussion, where the discussion is emerging from, myths shared about the virus, and how much of it is connected to other high and low quality information on the Internet through shared URL links. Our preliminary findings suggest that a meaningful spatio-temporal relationship exists between information flow and new cases of COVID-19, and while discussions about myths and links to poor quality information exist, their presence is less dominant than other crisis specific themes. This research is a first step toward understanding social media conversation about COVID-19. 
\end{abstract}


\section{Introduction}

In December 2019, a cluster of pneumonia cases caused by a novel coronavirus (COVID-19) was identified in Wuhan, China. In the months since, the outbreak has rapidly spread around the world, leading to hundreds of thousands of cases in over 160 countries~\cite{WorldHealthOrganization.}. This unparalleled global health emergency has resulted in an unprecedented political and social response, and is already having massive consequences on the global economy. The avalanche of human response is being facilitated by the flow of information from the broadcast world of traditional media but, in particular, by the networked world of social media.

Social media is a significant conduit for news and information in the modern media environment, with one in three people in the world engaging in social media, and two thirds of those on the Internet using it~\cite{OrtizOspina.2020}. The popularity is higher in the United States with 68\% of American adults reporting they get news on social media~\cite{Matsa.}. This is particularly true for health and science information, with a third of people reporting that social media are an “important” source of science news~\cite{Hitlin.12302019}. Twitter users, in particular, are known for sharing and consuming news: 59\% of Twitter users describe it as “good” or “extremely good” for sharing preventive health information~\cite{Wilford.2018}.

However, social media is also rife with health misinformation. Health misinformation - often defined as information that counters best available evidence from medical experts at the time (~\cite{Emily.2020} see also ~\cite{Garrett.2016,Nyhan.2010,Tan.2015,Southwell.2019}) - has been documented across almost all social media platforms, including Facebook, Twitter, YouTube, Pinterest, and Instagram~\cite{Briones.2012,Broniatowski.2018,Dredze.2016,Guidry.2015,Oyeyemi.2014,Sharma.2017}. Moreover, health misinformation is not limited to any one issue, but includes vaccination misinformation (both generally and specific cases such as HPV or flu), and misinformation about global health crises like the Ebola outbreak in 2014, and the spread of Zika in 2016. 

While past epidemics highlight the importance of studying infectious disease related social media content, there is a particular significance and urgency in studying social media content for COVID-19. First, we expect social media to play an even bigger role in the spread of information about COVID-19; for example, there were 255 million active Twitter users in February 2014 during the start of the Ebola outbreak~\cite{Twitter.April292014}. This number topped 330 million in 2019~\cite{Twitter.April232019}. Put simply, a lot more people are connecting to others online and getting their news through social media platforms such as Twitter~\cite{Matsa.,Hitlin.12302019}. Second, there is reason to be more concerned about the quality of such information in today's news ecosystem compared to that of earlier epidemics. As recent research shows, trust in institutions is eroding~\cite{Swift.2016} and this is accompanied by an uptick in the spread of misinformation online. Therefore, it is crucial to study and understand the conversation surrounding the fast-moving COVID-19 pandemic through the lens of social media. 

This article looks at the amount of conversation taking place on social media, specifically Twitter, with respect to COVID-19, the themes of discussion, where the discussion is emerging from, and how much of it is connected to other high and low quality information on the Internet through shared URL links. We also identify and look at the overall and temporal level of discussion of five specific myths shared on Twitter. We pause to mention that all of the signals we are measuring and using are, at best, moderate quality indicators from social media to reported cases of COVID-19. But it is precisely because we do not have access to high quality indicators that we must take the time to calibrate moderate and poor quality indicators, glean insight from these indicators, and understand their relationship to each other. While this paper will not tackle the measurement issues associated with indicators, we recognize its importance as part of our longer term research agenda.

Our preliminary findings suggest that (1) conversation about the virus continues to grow, (2) for some countries, information flow leads new cases of COVID-19 by 2-5 days, (3) predominant themes of conversation include health/the virus itself or the global nature of the pandemic, and (4) misinformation and myths are discussed, but at lower volume than other conversation. While many of these findings are not surprising, they remind us of the importance of social media during times of crisis and give preliminary support for using this open platform as a surveillance approach for understanding how people are impacted by a crisis.

\section{Size of COVID-19 Conversation}

Conversation size gives us insight into the amount of attention placed on a topic. In this section we measure the amount of conversation taking place about COVID-19 in general and by language.

\subsection{Description of Data Set}
Using the Twitter Streaming API, we began collecting tweets related to COVID-19 on January 16, 2020. Data collection continues, but the data we present in this study is from January 16, 2020 to March 15, 2020. Table~\ref{table-hastags} in Appendix A shows the English hashtags we used to collect data and the dates we began collecting data for the hashtag. 
Most of the data collection began in January, and additional hashtags were added in mid-March to reflect the changing nature of the conversation around COVID-19 online.\footnote{Note that due to a data collection glitch with the Twitter Streaming API, some of the hashtags were unavailable between March 13, 2020 and March 15, 2020.}

\subsection{General Conversation Volume}
\begin{figure}[!htb]
\includegraphics[width=0.9\linewidth]{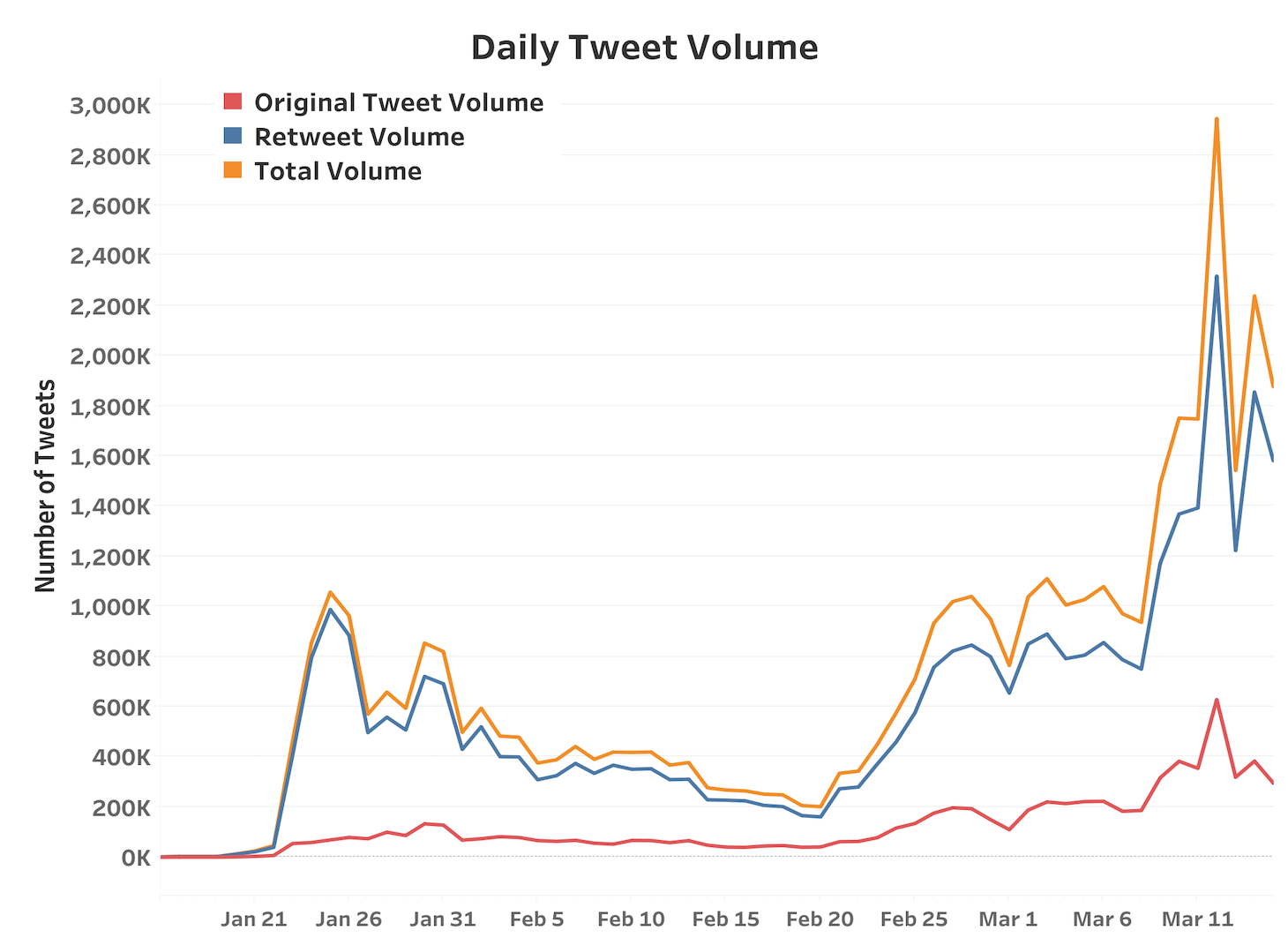}
\caption{Overall Volume of Tweets (1/16/20 -- 3/15/20)}
\label{fig:tot_tweet_vol}
\end{figure}

During the study period of January 16 through March 15, the overall number of tweets is 2,792,513, quotes are 456,878, and retweets are 18,168,161. Figure \ref{fig:tot_tweet_vol} shows the overall volume of tweets, and the volume of tweets and retweets separately. Initial tweets about COVID-19 were relatively infrequent. It is not surprising that both the tweet and retweet volume continue to increase as the epidemic unfolds. What is interesting is that there is no single event that has propelled the increase. The peak volume days map to a series of events unfolding throughout the crisis.  

There was a significant ramping up in late January around the 25th. On January 24th, multiple countries, including Japan, South Korea, and the United States, reported cases, and the increasing number of cases in China led the Chinese government to begin a quarantine in the Hubei province~\cite{cnn.01242020}.  On January 25th, Hong Kong declared a state of emergency and the United States also announced plans to evacuate US citizens from Wuhan~\cite{cbs.02062020}. This was followed by a tapering off in the first half of February. During this time, the cases continued to grow around the world, but little government action was taking place outside of Asia. At the end of February, another large increase in volume occurred. That was when the number of deaths in Iran and Italy grew over 30 and 20, respectively, and Switzerland banned all gatherings/events larger than 1000 people. February 29th was also the date of the first death due to COVID-19 in the United States~\cite{Holland.02292020}. Following that time period, there was an overall increasing trend, with volume essentially doubling from the first to the second week in March.\footnote{Again, recall that numbers for March 13 and 14 are lower due to an API glitch, so the significant dropoff at that point is merely an artifact} Essentially, this suggests that attention to COVID-19 has increased significantly over the last two months, and is growing at a faster rate in March than in January. 

\begin{figure}[!bt]
\centering
\includegraphics[width=0.99\linewidth]{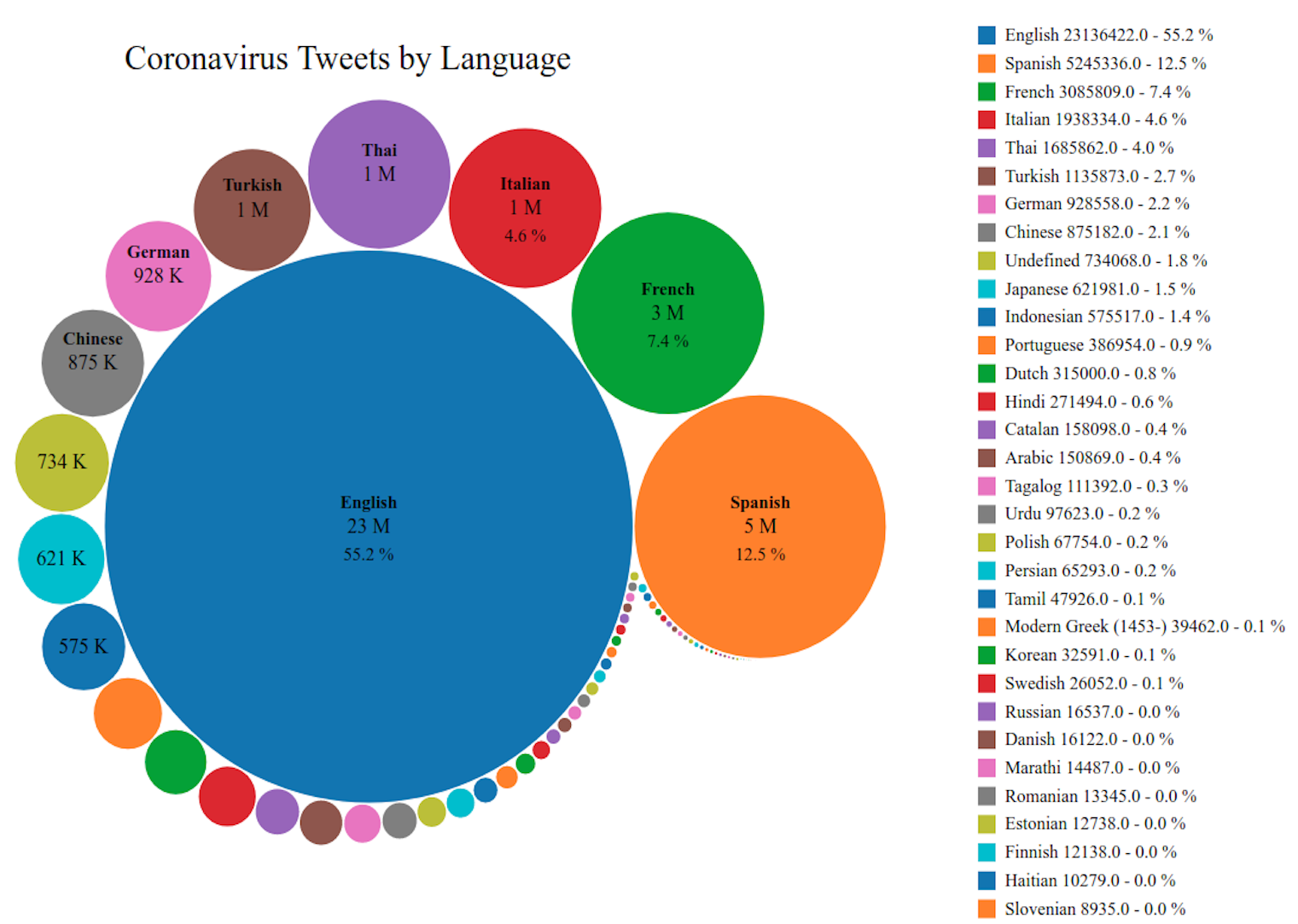}
\caption{COVID-19 Tweets by Language (1/16/20 -- 3/15/20)}
\label{fig:tot_tweet_lang_bubble}
\end{figure}%

\begin{figure}[t]
\centering
\includegraphics[width=0.7\linewidth]{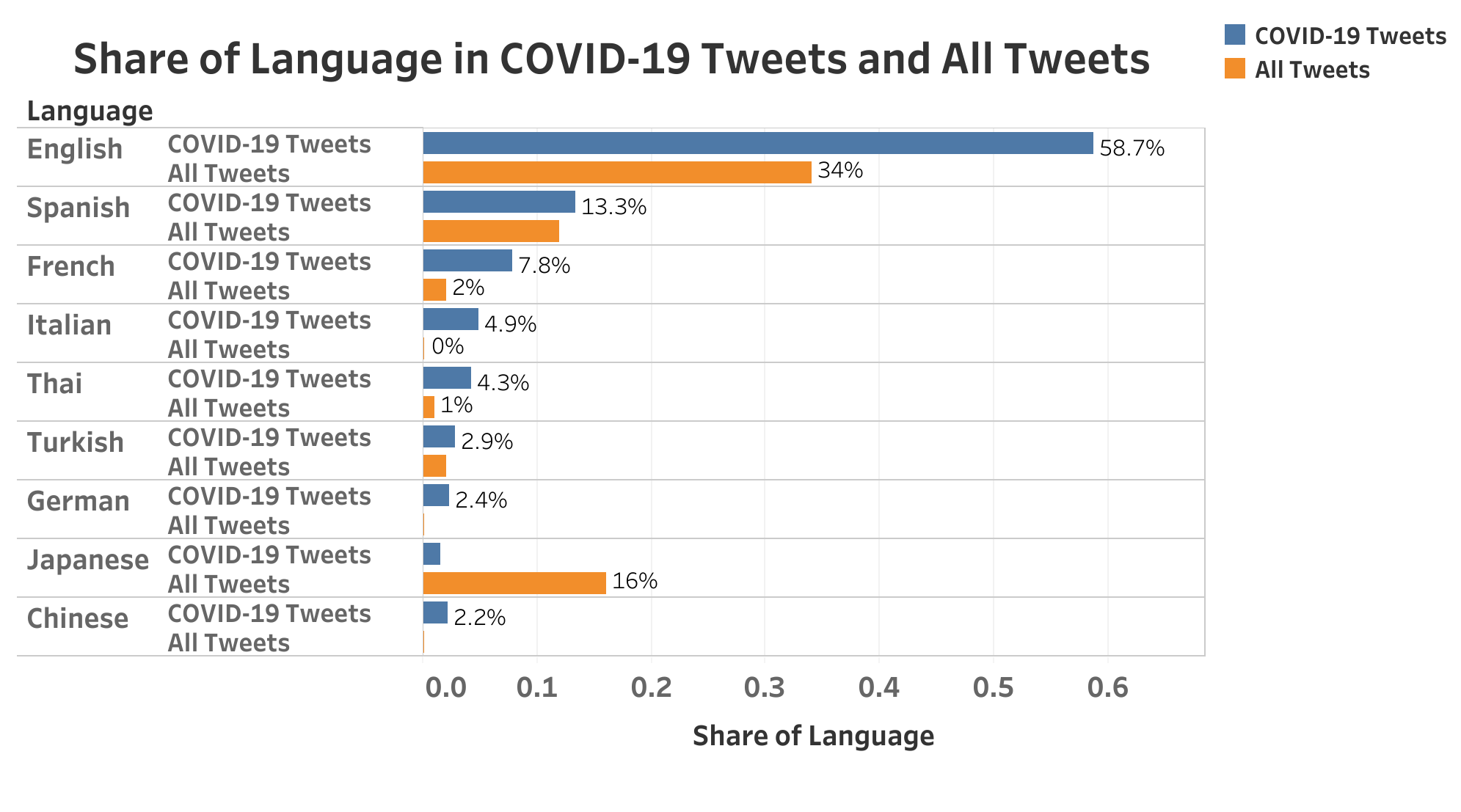}
\caption{Relationship Between COVID-19 Tweet Languages and Overall Twitter Language Distribution (1/16/20 -- 3/15/20)}
\label{fig:tweet_lang_comp}
\end{figure}

\subsection{Tweet Volume By Language}
Twitter specifies the language of each tweet when using its streaming API. To get a better sense of the global nature of the conversation, we look at the proportion of tweets in different languages over time in Figure \ref{fig:tot_tweet_lang_bubble}. Not surprisingly, given worldwide Twitter usage, the majority of tweets (57.1\%) are in English. This is followed by Spanish (11.6\%), French (6.5\%), and Italian (4.8\%). While English and Spanish are generally dominant languages on Twitter (see Figure \ref{fig:tweet_lang_comp}), we see that the language distribution for COVID-19 tweets does vary from the overall Twitter population~\cite{Orcutt.2013} for English and other languages. The languages that are most prevalent are the predominant languages in countries where many of the outbreaks have taken place during this time, including the United States (1,678 confirmed cases as of March 15), Spain (5,753 confirmed cases), France (4,469 confirmed cases), and Italy (21,157 confirmed cases)~\cite{WorldHealthOrganization.}. Given their significant outbreaks, we might have expected more tweets in Chinese (81,048 confirmed cases) and German (3,795 confirmed cases), although Twitter use in both countries is relatively low (Twitter is blocked in China, although some people use VPN to get around that limitation; Twitter is used by only 9.8\% of adults in Germany~\cite{StatcounterGlobalStats.2020}). The languages that are more prevalent on Twitter but not as represented in the COVID-19 tweets when considering the overall distribution of language on Twitter are Indonesian, Arabic, and Malay.

\begin{figure}
    \centering
    \caption{Tweet Volume By Language}
    \label{fig:vol_lang}
        \subfloat[Chinese and English]{\includegraphics[width=0.65\columnwidth]{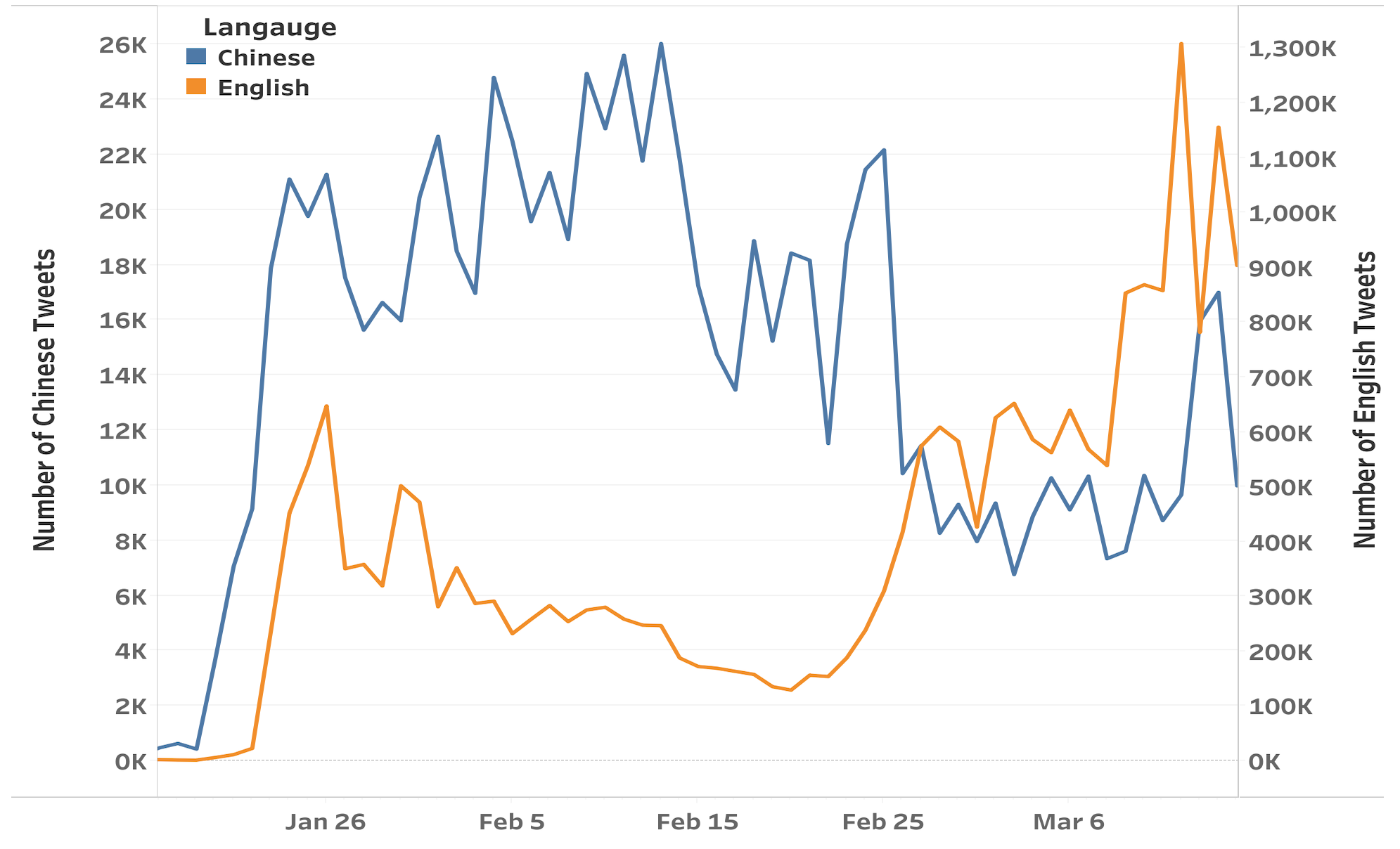}\label{fig:chinese_eng_tweet_vol}} \\ \vspace{3mm} \qquad
        \subfloat[French, German, Italian, Spanish and Turkish]{\includegraphics[width=0.65\columnwidth]{fig/chinese_eng_tweet_vol_v2.png}\label{fig:other_lang_tweet_vol}} \vspace{3mm}  \\
        \qquad
        \subfloat[Japanese and Thai]{\includegraphics[width=0.65\columnwidth]{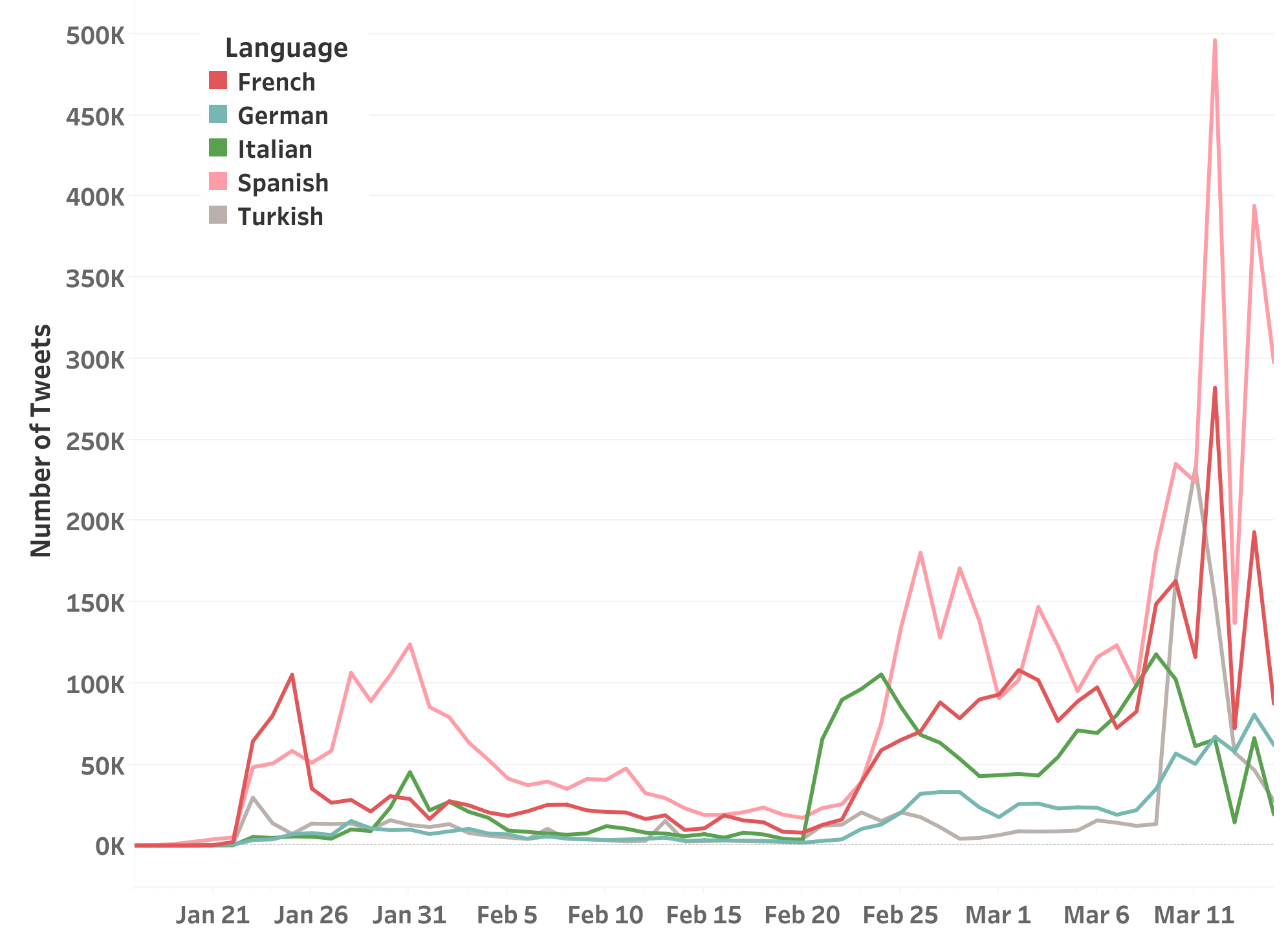}\label{fig:jp_thai_tweet_vol}}
\end{figure}

Focusing in on the languages that are more prevalent in Figures \ref{fig:tot_tweet_lang_bubble} and \ref{fig:tweet_lang_comp}, we look at how the volume changes from mid-January to mid-March in Figure \ref{fig:vol_lang}. Figure \ref{fig:chinese_eng_tweet_vol} shows the higher volume of tweets in January and early February in Chinese, and a contrasting view of volume for the tweets in English. For English, there is a small spike in January, with a decrease in volume until late February. This is when more cases began emerging in North America and Europe. The trend that English follows is similar for French, German, Italian, Spanish and Turkish - an initial increase followed by a period of low discussion followed by a large increase and spikes in March. This can be seen in Figure \ref{fig:other_lang_tweet_vol}. Spanish, being spoken significantly in North America and Europe and having a large Twitter user base has the highest increase. Finally, Figure \ref{fig:jp_thai_tweet_vol} shows a contrast in two predominantly Asian Languages, Japanese and Thai. After an initial spike in January when initial cases were confirmed in both these countries, the volume of Japanese conversation has been fairly constant. In contrast, the volume of Thai tweets had a very high spike in January and a high one in March. This difference in trend may be related to doubling in the number of cases reported in Thailand between late February and min-March and the first COVID-19 death in March \cite{Wongchaum.}. 

\section{Location of Conversation and Its Relationship to COVID-19 Cases}

Previous work has shown that social media can be used to help inform where people may move during periods of forced migration \cite{Singh.2020}. In this section, we are interested in extending this idea to see if social media can help inform the public about movement of a disease. To begin to understand the relationship between these different signals, we look at the relationship between where people tweet from, what locations people tweet about, and reported COVID-19 cases. 

\subsection{Description of Data Sets}
We use two sets of Twitter data to look into this. The first set is extracted from our Twitter data with location mentions; for example, if both China and COVID-19 are mentioned in a tweet, that tweet will be assigned to China regardless of where that tweet is sent from. We refer to this set of tweets as conversation location tweets. To determine the locations mentioned in the tweet, we compare the words in the tweet to a location ontology we created using Wikipedia and Statoids~\cite{Statoids.2016} that contains cities, provinces, states, and countries.\footnote{Wikipedia has a set of pages listing  the major cities around the world by country. Statoids lists governorates and the governorates’ capitals for each country. Combining these two sources, we construct an ontology containing approximately 7,600 locations, including countries, governorates, governorates’ capitals, and other major cities.}

Our second data set contains geotagged tweets, which are tweets that users choose to tag with a longitude and latitude location (this is relatively rare, as it is not the default setting in Twitter). This allows us to track with greater confidence where the tweets come from. We were able to collect 100\% of the existing geotagged tweets mentioning any of our hashtags through the Twitter API. 
Our final data set comes from the World Health Organization. It contains the number of confirmed COVID-19 cases in different countries as of March 15, 2020~\cite{WorldHealthOrganization.}. 

\subsection{Locations of Twitter conversation}
\begin{figure}
    \centering
    \caption{Tweet Volume and COVID-19 Cases }
    \label{fig:volume_covid_map}
        \subfloat[Conversation Location]{\includegraphics[width=0.9\columnwidth]{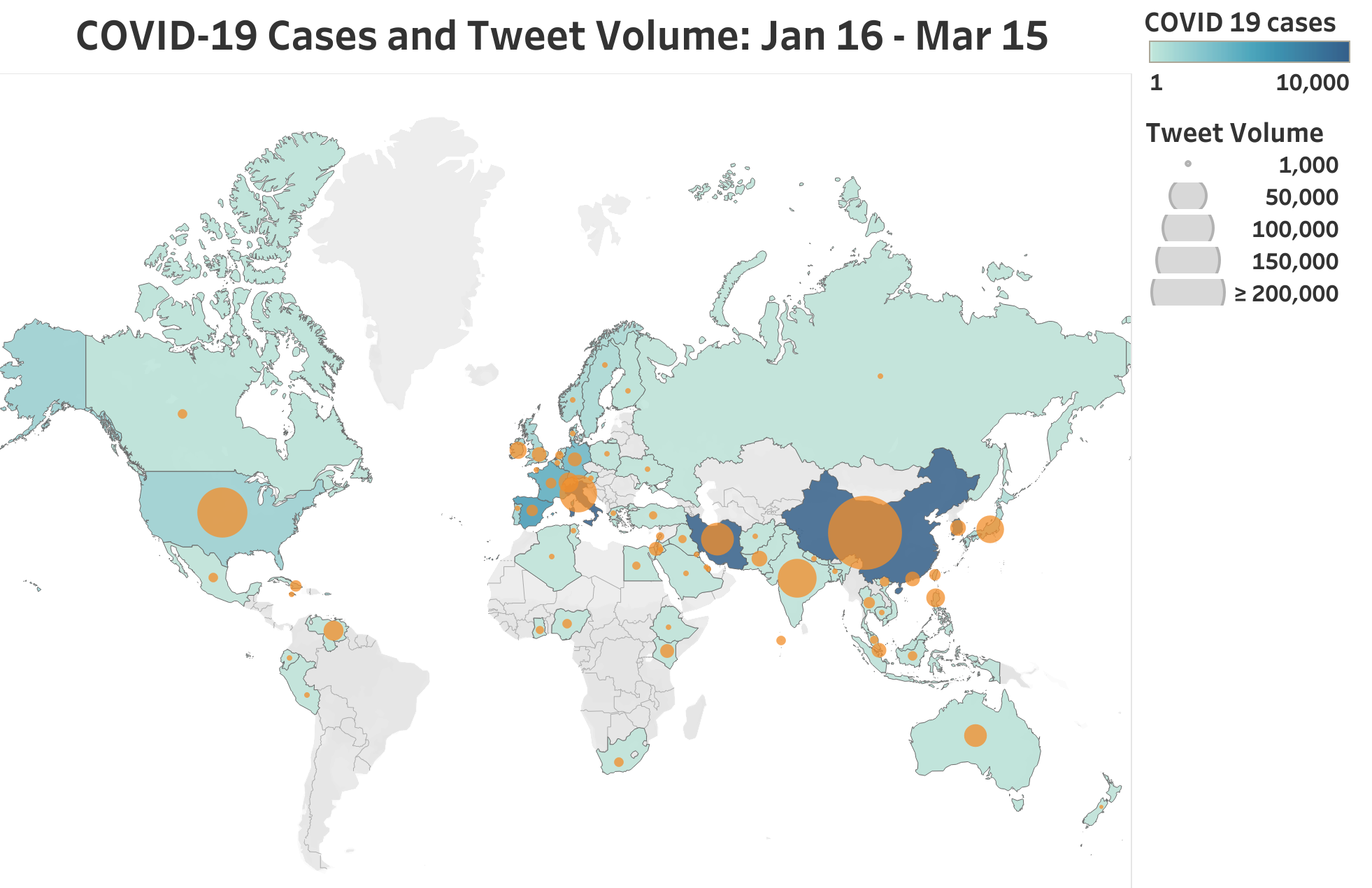}\label{fig:conversation_map}} \\ \vspace{3mm} \qquad
        \subfloat[Geo-tagged Location]{\includegraphics[width=0.9\columnwidth]{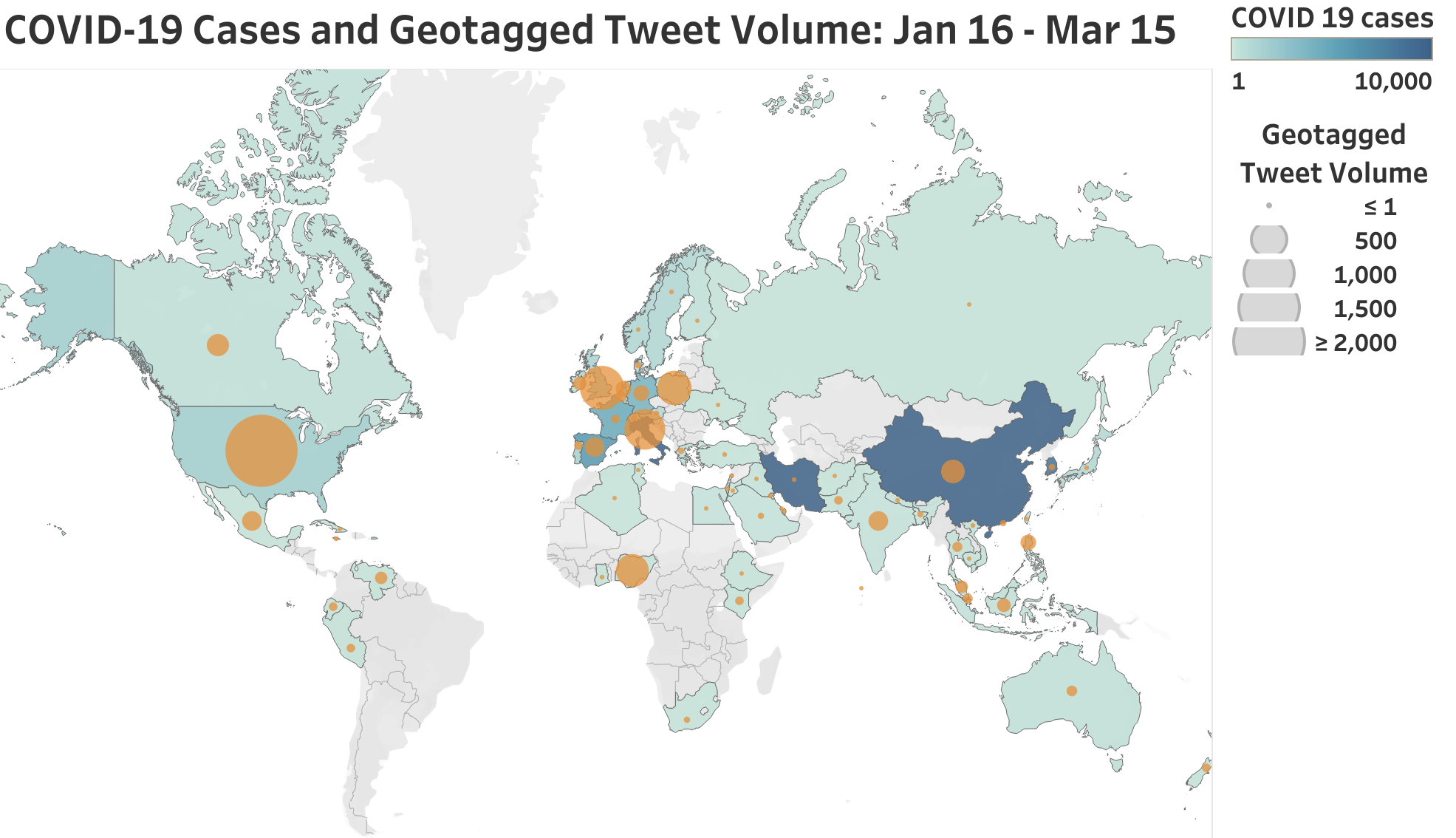}\label{fig:geotag_map}}
\end{figure}

Figure \ref{fig:volume_covid_map} shows two maps each containing the COVID-19 reported case count and one of our Twitter location data sets. 
Figure \ref{fig:conversation_map} illustrates the number of confirmed COVID-19 cases and the number of conversation location tweets by country. 217 countries were mentioned and 184 of them were mentioned at least 10 times. Overall, it appears that conversation location tweets and confirmed cases of COVID-19 are highly correlated; countries that have more cases also have more conversation location tweets (tweets that mention the country itself or a location in that country). Among the conversation location tweets, COVID-19 was mentioned the most in China, followed by the USA, India, Iran, and Italy. It makes sense that China has the most COVID-19 mentions from January 16 to March 15, when China was the epicenter of COVID-19. The USA has the second highest tweet mentions, although the USA ranks only 8th place in terms of cases by March 15; this is likely because the USA has the largest Twitter user population and the most Asian immigrants, who may have been following COVID-19 trends earlier than other population groups.

Figure \ref{fig:geotag_map} maps the number of geotagged tweet mentions of COVID-19 by country. Note that these tweets were sent out from the corresponding countries and that so far we have collected geotagged tweets in English only. The United States has the most geotagged tweet mentions of COVID-19, followed by the United Kingdom, Italy, Poland, Nigeria, China, Spain, and India. It is not surprising that the United States and the United Kingdom have the most geotagged tweet mentions since both are English speaking countries and our geotagged tweets are limited to English only. During the study period, Italy has the second most COVID-19 cases and the third most geotagged tweet mentions. It is interesting that although Twitter is restricted in China and most Chinese people speak Chinese only, China has the sixth most geotagged tweet mentions from January 16 to March 15, when China was the epicenter of the COVID-19. When we compute the Pearson correlation between the volumes of the location conversation tweets and the geotagged tweets by countries, we get a correlation of 0.75, if we exclude China from the calculation. While not a perfect proxy for geotagged tweets, it is a reasonable one.

\subsection{Comparison between location conversation and COVID-19}
We take a deeper dive into three countries (the United States, Italy, and China), chosen because 1) they each have a large number of confirmed COVID-19 cases, 2) they are each on a different continent, offering some variance in location and culture, and 3) they are at different stages of their epidemic trajectory, allowing some insight into how conversations might change over the course of an outbreak in a country. 

\begin{table}[h]
\centering
\caption{Pearson Correlations between Location Conversation and COVID Cases with Different Leads and Lags in the USA, Italy, and China.}
\label{table-pearson-correlations}
\begin{tabular}{|l|c|c|c|c|}
\hline
\rowcolor[HTML]{CBCEFB} 
\multicolumn{1}{|c|}{\cellcolor[HTML]{CBCEFB}\textbf{Lag/Lead}} & \textbf{US} & \textbf{Italy} & \textbf{China-A} & \textbf{China-B} \\ \hline
Lag = 5 days   & 0.492          & 0.645          & 0.037          & 0.295          \\ \hline
Lag = 4 days   & 0.487          & 0.649          & 0.141          & 0.371          \\ \hline
Lag = 3 days   & 0.458          & 0.698          & 0.156          & 0.488          \\ \hline
Lag = 2 days   & 0.617          & 0.748          & 0.208          & 0.565          \\ \hline
Lag = 1 day    & 0.667          & 0.724          & 0.275          & 0.610          \\ \hline
No lag or lead & 0.639          & 0.744          & 0.293          & 0.680          \\ \hline
Lead = 1 day   & 0.693          & 0.699          & 0.283          & 0.720          \\ \hline
Lead = 2 days  & \textbf{0.780} & 0.709          & 0.291          & 0.753          \\ \hline
Lead = 3 days  & 0.645          & 0.837          & 0.344          & 0.771          \\ \hline
Lead = 4 days  & 0.637          & \textbf{0.880} & \textbf{0.425} & \textbf{0.794} \\ \hline
Lead = 5 days  & 0.649          & \textbf{0.885} & 0.391          & \textbf{0.803} \\ \hline
\end{tabular}%
\tabnote{\textsuperscript{a}All correlations for the United States, Italy, and China-B are statistically significant at p <= 0.05.}
\end{table}

\begin{figure}
    \centering
    \caption{The number of COVID-19 Cases and Location Conversation}
    \label{fig:vol_lang}
        \subfloat[United States: Lead = 2 Days]{\includegraphics[width=0.7\columnwidth]{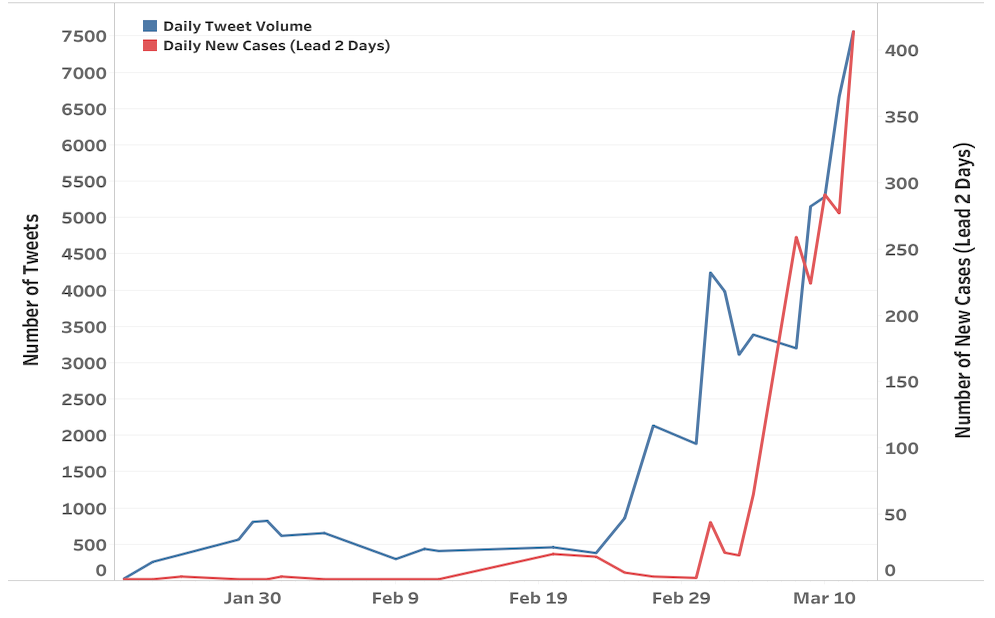}\label{fig:usa_case_lead}} \\ \vspace{3mm} \qquad
        \subfloat[Italy: Lead = 4 Days]{\includegraphics[width=0.7\columnwidth]{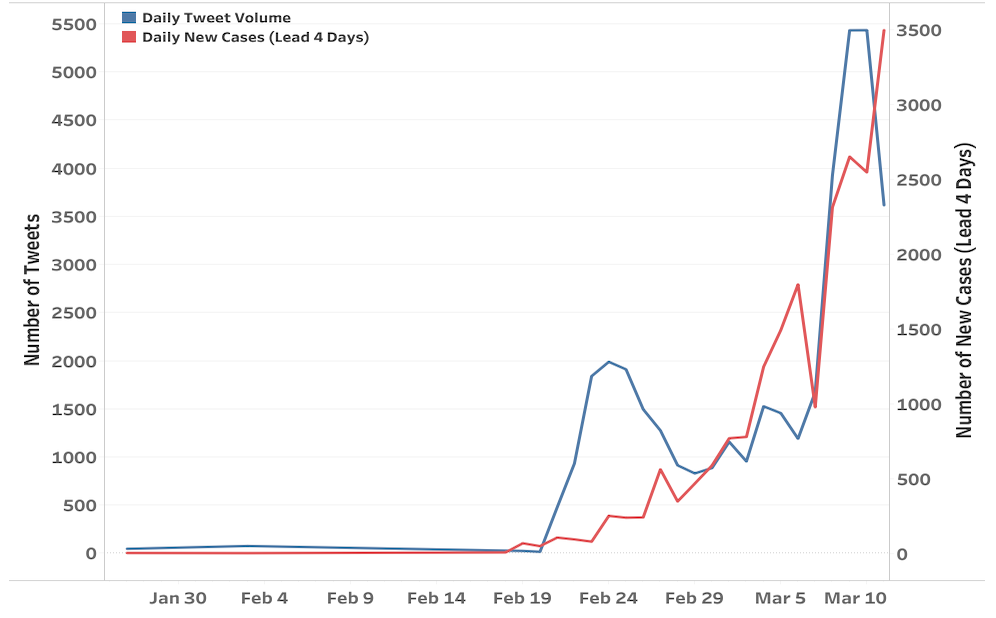}\label{fig:italy_case_lead}} \vspace{3mm}  \\
        \qquad
        \subfloat[China: Lead = 4 Days]{\includegraphics[width=0.7\columnwidth]{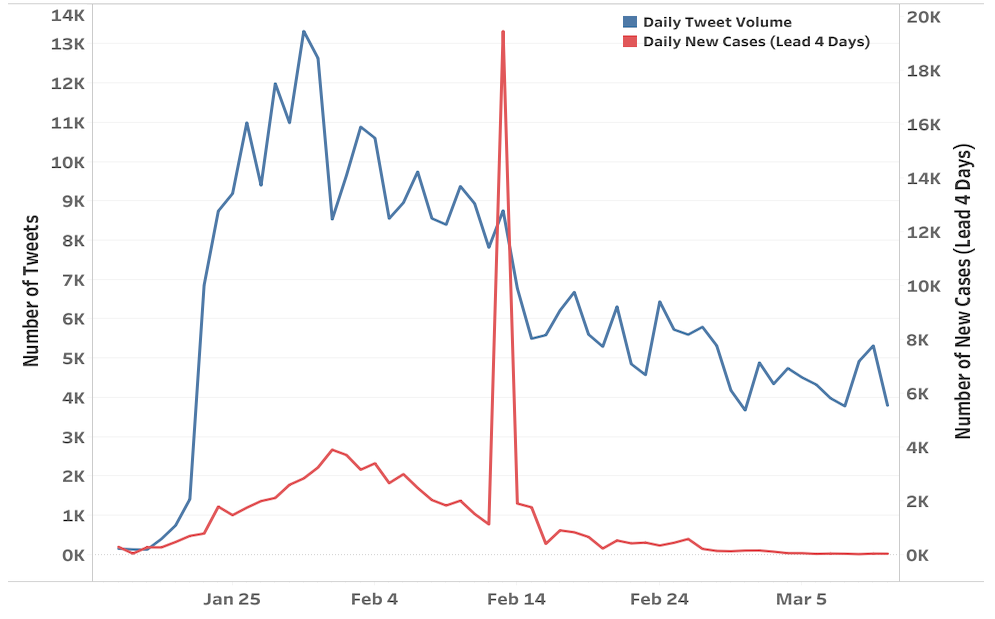}\label{fig:china_case_lead}}
\end{figure}

This initial analysis focuses on the time series of the location conversation tweets and the newly identified and confirmed COVID-19 cases each day. We compute the Pearson correlation between time series of location conversation tweets and COVID-19 cases with different leads and lags to determine if a relationship exists between the two variables. We hypothesize that social media communications may be predictive of cases, creating opportunities for disease forecasting. 

The cross-correlation analysis  between location conversation and COVID cases is presented in Table~\ref{table-pearson-correlations}. For China we have two columns, one with all the dates (China-A) and one that removes the single day spike on February 12, 2020 due to a change in testing procedure (China-B)~\cite{Gunia.02132020}. The numbers in bold are the highest correlations for each country. If the highest and second highest correlation values are within 0.01 of each other, we highlight both. A \textit{lead} refers to the tweets occurring before the cases; for example, a two-day lead means that we match the tweet conversations with the number of new cases two days later. In contrast, a \textit{lag} refers to the tweets occurring after the cases; a two-day lag means that we match the tweet conversations with the number of new cases from two days earlier.

We find that social media conversations are more highly correlated with COVID-19 cases with a lead than with a lag (Table~\ref{table-pearson-correlations}). That is, it seems that tweets increase in volume before confirmed cases increase in a given country. This suggests that social media conversations may be a leading indicator of disease cases, which can be explained by the lag between symptom onset and the progression of severe symptoms leading to testing (though this might depend on different protocols for testing implemented in different countries). Additionally, we find that conversations mentioning COVID-19 and locations in the United States, Italy and China-B all show a strong association to confirmed cases. This result suggests that the association between conversations and cases is indicative of both a direct effect and an indirect effect given the lack of Twitter usage in China. We also find that the lead time of 2 days for the US (Figure \ref{fig:usa_case_lead}), 4-5 days for Italy (Figure \ref{fig:italy_case_lead}), and 4-5 days for China-B (Figure \ref{fig:china_case_lead}) demonstrate the strongest correlations between location conversation mentions and confirmed cases.

\section{Content of English Conversation}
What are the most prevalent words and themes of discussion taking place about COVID-19? In this section, we take a first look at the content of the conversation taking place on Twitter. 

\subsection{Words Being Used}
\begin{figure}
    \centering
    \includegraphics[width=0.75\linewidth]{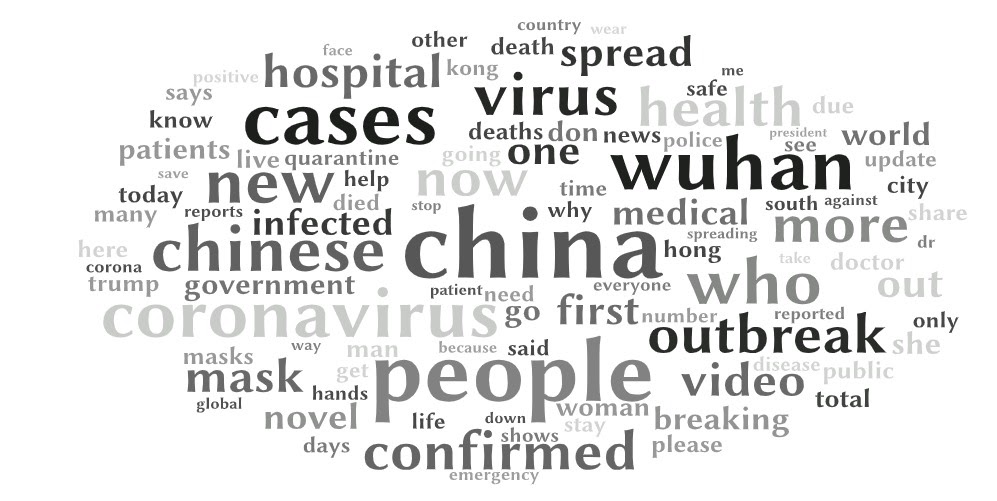}
    \caption{Word Cloud of Frequently Mentioned Words in COVID-19 Tweets
}
    \label{fig:word_cloud}
\end{figure}

We begin by looking at the most frequent words in each tweet.  Table~\ref{table-top-words-twitter} shows the top 10 words, excluding stopwords, and the number of tweets they appear in. It is not surprising that most of these words are very broad. To expand on this, Figure \ref{fig:word_cloud} shows a word cloud containing words that appear in at least 150,000 tweets. Words are sized based on how frequently they appear with larger words appearing in more tweets. We see that the dominant words across our data set focus on the global nature of the virus, words describing the virus and its spread, and responses to the outbreak. A strong focus on China reflects the focus of attention during the early phases of this epidemic.

\begin{table}[b]
\centering
\caption{Top 10 Most Frequently Used Words in Tweets.}
\label{table-top-words-twitter}
\begin{tabular}{|l|c|}
\hline
\rowcolor[HTML]{CBCEFB} 
\multicolumn{1}{|c|}{\cellcolor[HTML]{CBCEFB}\textbf{Word}} & \textbf{Volume} \\ \hline
china & 1,412,521 \\ \hline
people & 1,155,989 \\ \hline
cases & 929,914 \\ \hline
wuhan & 841,901 \\ \hline
coronavirus & 821,484 \\ \hline
new & 793,917 \\ \hline
chinese & 765,332 \\ \hline
who & 712,041 \\ \hline
virus & 615,342 \\ \hline
confirmed & 585,393 \\ \hline
\end{tabular}
\end{table}

\subsection{Prevalent Themes}
Since words alone give us limited insight into people’s themes/topics of conversation, we next identify common themes in tweets about COVID-19 and how the prevalence of these themes change over time. 

\subsubsection{Methodology}
To achieve this, we first identify the set of frequent words that are observed at least 50,000 times in our tweets data set. This list includes 537 unique words. These words are then grouped into themes by three researchers through open coding. We identify eight high level categories: Economy, Emotion, Illness, Global Nature, Information Providers, Social, Government Response, and Individual Response. We provide details about these topics below in Table~\ref{table-summary-of-themes}. After determining the set of themes, word-to-theme mapping was constructed through majority voting--a word is assigned to a given topic if at least two out of the three researchers agreed that the assignment was correct. The words that do not belong in any of these categories (e.g. stop words) are removed. Word variations were also added (e.g. plurals) to increase coverage. While these themes are a reasonable first pass, we plan to improve these in the future using semi-supervised, iterative topic modeling since fully automated topic modeling is less effective on tweets.

For each tweet, we identify the words that map from each theme and proportionally assign themes to the tweet. Then the proportions are summed together per day to determine the overall tweet volume of the different themes. Approximately 80\% percent of the tweets were labeled with one or more themes.

\begin{table}[h]
\centering
\caption{Summary of the themes identified using most frequent words in tweets.}
\label{table-summary-of-themes}
\resizebox{\textwidth}{!}{%
\begin{tabular}{|l|c|l|}
\hline
\rowcolor[HTML]{CBCEFB} 
\multicolumn{1}{|c|}{\cellcolor[HTML]{CBCEFB}\textbf{Theme}} & \textbf{\begin{tabular}[c]{@{}c@{}}Number\\ of Words\end{tabular}} & \multicolumn{1}{c|}{\cellcolor[HTML]{CBCEFB}\textbf{Example words}} \\ \hline
Economy & 12 & Market, stocks, futures \\ \hline
Emotion & 24 & Fear, joke, hope \\ \hline
Healthcare/Illness/Virus & 64 & Patients, coronavirus, infected, vaccine, tested, sars \\ \hline
Global Nature & 75 & Pandemic, international, China, Italy, travel \\ \hline
Information Providers & 28 & Media, CDC, WHO, experts \\ \hline
Social & 6 & Family, friends, community \\ \hline
Government/ Government Response & 28 & Trump, senator, lockdown \\ \hline
Individual Concerns/Strategies & 28 & Disinfect, wash, facemasks \\ \hline
\end{tabular}%
}
\end{table}

\subsubsection{Findings}
\begin{figure}
\centering
\includegraphics[width=0.7\linewidth]{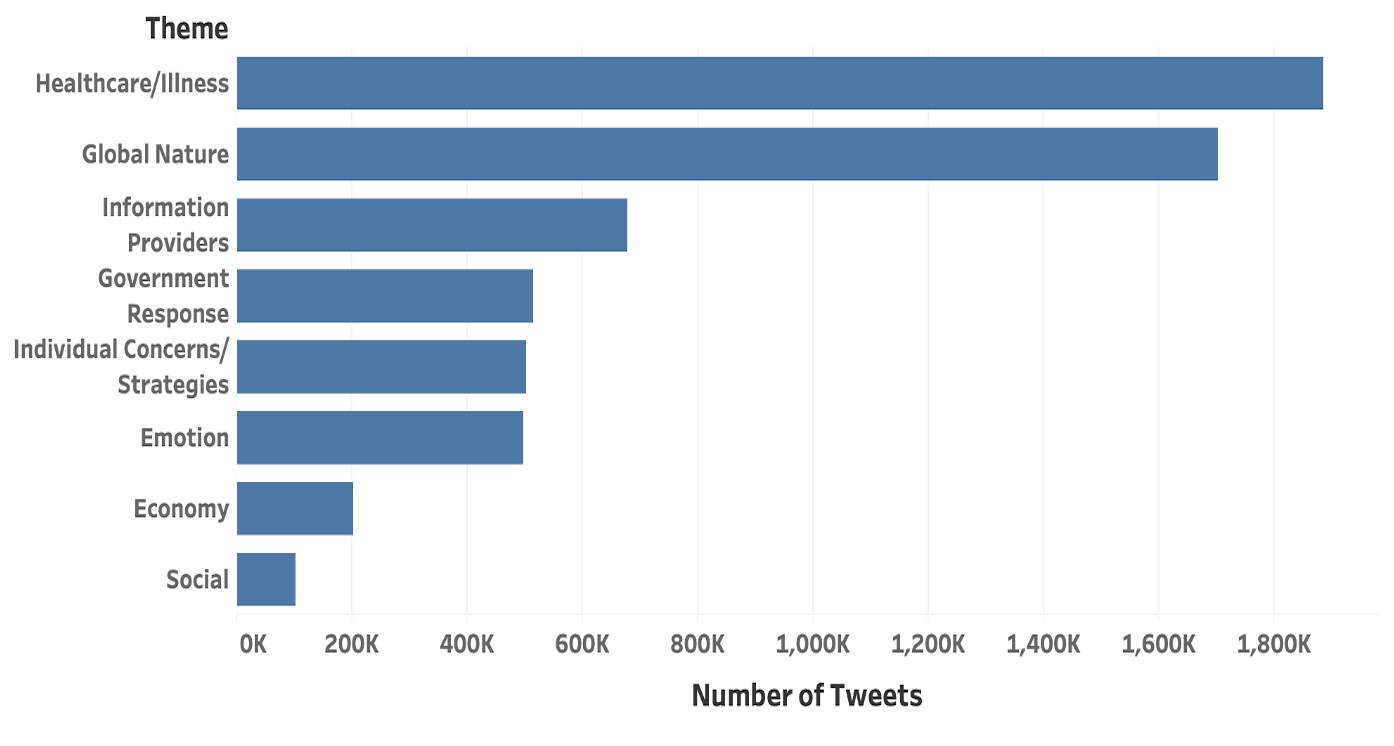}
\caption{Distribution of Themes in Tweets}
\label{fig:theme_bar}
\end{figure}

Figure \ref{fig:theme_bar} summarizes the overall distribution of the themes presented in Table~\ref{table-summary-of-themes}. We observe that most Twitter conversations are about one of two topics: either health/the virus itself or the global nature of the pandemic. It is not surprising since these categories contain a larger number of broader terms about the pandemic. The first category includes conversations about the virus itself, broader health consequences, vaccines, testing, and references to other epidemics. This accounts for 30\% of the labeled content. The second similarly-sized theme is the global nature of COVID-19. Twitter users are commonly referring to locations around the world where the disease is spreading (e.g. China, Italy) and referring to its scale (e.g. pandemic, worldwide). This theme accounts for 29\% of labeled content. The next biggest theme is information providers (11\%). This suggests that a sizable number of tweets are referring to or talking about sources of information about the disease (e.g. CDC or news media).  In the time period we consider, the economy theme was rather rare (3\%). This fact might change over time if disease control improves and the immediate effects of the disease are overshadowed by secondary effects, including possible economic consequences. Finally, we note that the emotion theme, containing words like fear, sad, hope, makes up 9\% of the labeled tweets. This is a significant fraction, and a reminder that attention needs to be given to mental health needs during this crisis.

\begin{figure}
\centering
\includegraphics[width=0.7\linewidth]{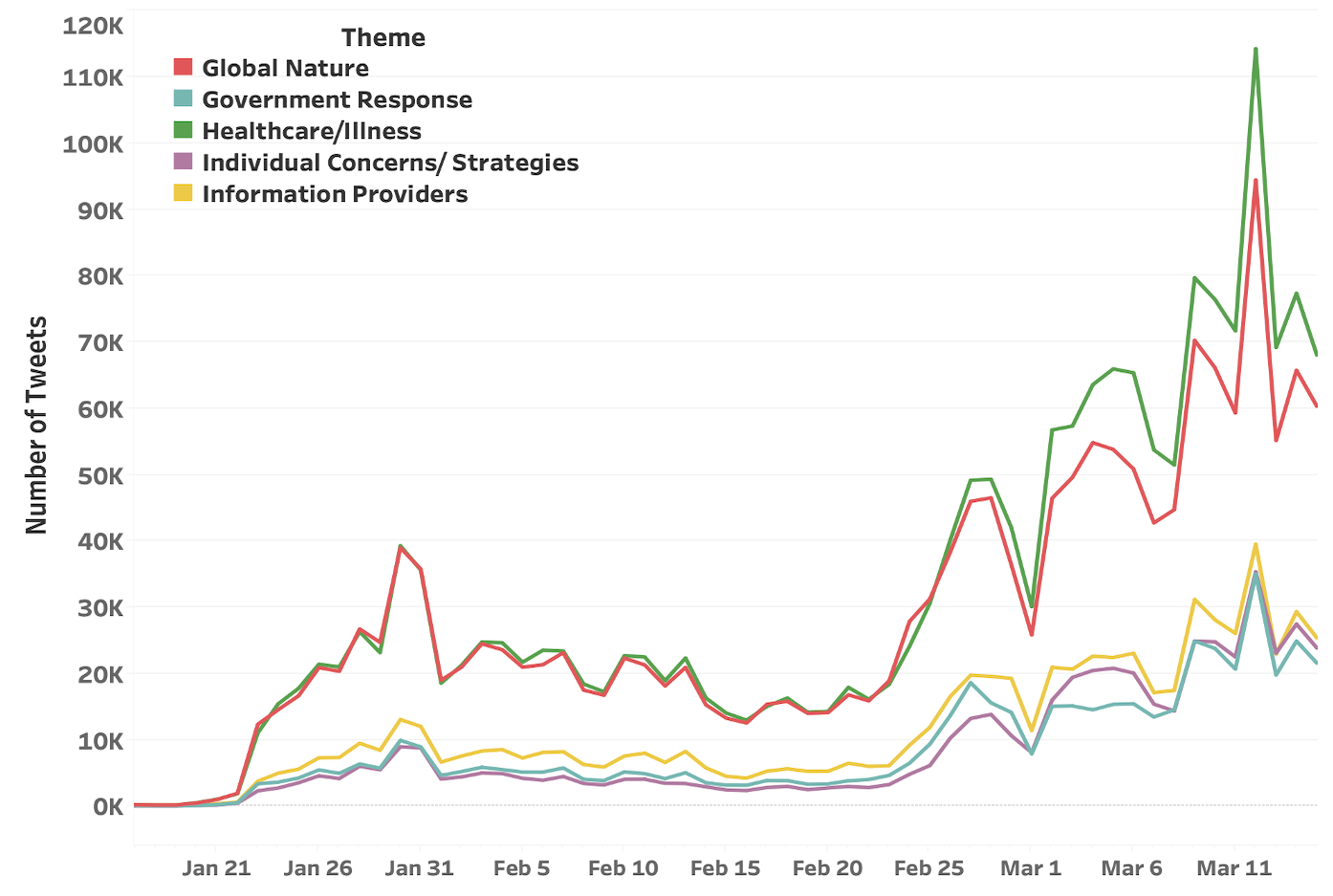}
\caption{Volume of Top 5 Themes in Tweets over Time}
\label{fig:theme_trend}
\end{figure}

As the epidemic grows and evolves, it is plausible that the themes and their prevalence will also evolve. Therefore, we next inspect the time series of the prevalence of the top five themes. The results are summarized in Figure \ref{fig:theme_trend}. We observe that all the themes have a similar trend to the trend of the overall volume for COVID-19 hashtags. The global nature of the disease and the health/virus themes are the most common theme early on. This is likely due to tweets referring to China when talking about COVID-19 and general conversation about the virus and cases. Starting with the last week of February, as the reality sets in that COVID-19 is spreading in the United States, all the themes become more popular, highlighting the continued diversity of the conversation.

\section{Myths About the Virus}
A large number of myths have emerged during this crisis. In this analysis, we are interested in determining how many tweets are discussing some of the most prevalent myths. 

\subsection{Methodology}
To determine what myths to include in our analysis, we searched different websites using the search phrase “Coronavirus Common Myths”. Then we used the articles to determine the most common examples. We analyzed a variety of sources including blogs, newspapers/media, and medical organizations to get a variety of examples. We then identified common themes among the different myths; for example, grouping weather and heat myths into a single category. From there, we manually chose the final list of myths based on how frequently they appeared during the search process and how dangerous they were. For example, while rumors about certain celebrities testing positive were common, they were not as dangerous to the public as myths about treatments or vaccines. We initially identified ten myths ranging from home remedies to conspiracy theories about the origin of the virus to misinformation about warm weather killing the virus.

To understand how certain myths about COVID-19 gained traction on social media, we searched for each of these myths in our tweet collection. We preprocessed the tweets to normalize capitalization and remove punctuation, and URLs. For each common myth, our team identified ten tweets that perpetuated the relevant myth. We identified phrases and words from the tweets and broad descriptions of the myths from our original Google searches that described each myth. For each such word/phrase, we assign a weight to signify the expectation that a tweet including this word/phrase would be about that given myth. For instance, the word “bioweapon” is a stronger signal about a \texttt{bioweapon} myth compared to \texttt{government lab} which occurs in \textit{fewer} bioweapon myth tweets. 

\begin{figure}
\centering
\includegraphics[width=0.6\linewidth]{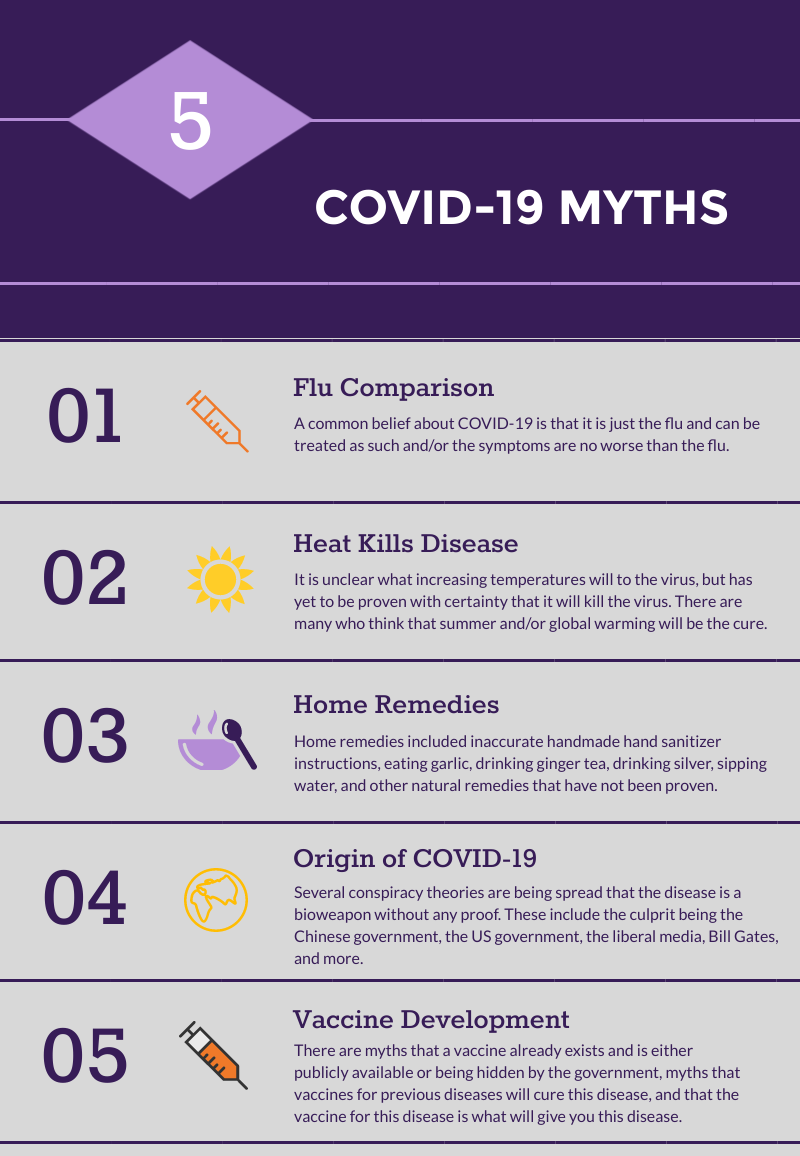}
    \caption{Myths about COVID-19}
    \label{fig:myth_list}
\end{figure}

After completing the word/phrase list for each myth, we then searched for words and phrases from each myth in each tweet, and proportionally assigned myths to tweets. The proportions are then summed together per day to determine the tweet volume of each myth. The results presented focus on the five myths that we identify most accurately: Origin of COVID-19, Vaccine Development, Flu Comparison, Heat Kills Disease, Home Remedies. Figure \ref{fig:myth_list} presents a description of each of these five myths.\footnote{Given the simple approach, we recognize that some tweets with more noise will be missed. However, we are erroring on the side of being more conservative, i.e. optimizing for precision instead of recall. After applying this approach, we manually validated 250 tweets, 25 for each category, and 25 that were identified as non-myths. Five of our categories had an overall test precision of 80\% and our precision for identifying non-myths in our test set was 100\%. In other words, 80\% of the hand-validated tweets belonged to the myths, and 100\% of the tweets coded as not matching the myths did not contain the myths. }

\subsection{Findings}
\begin{figure}
    \centering
    \includegraphics[width=0.7\linewidth]{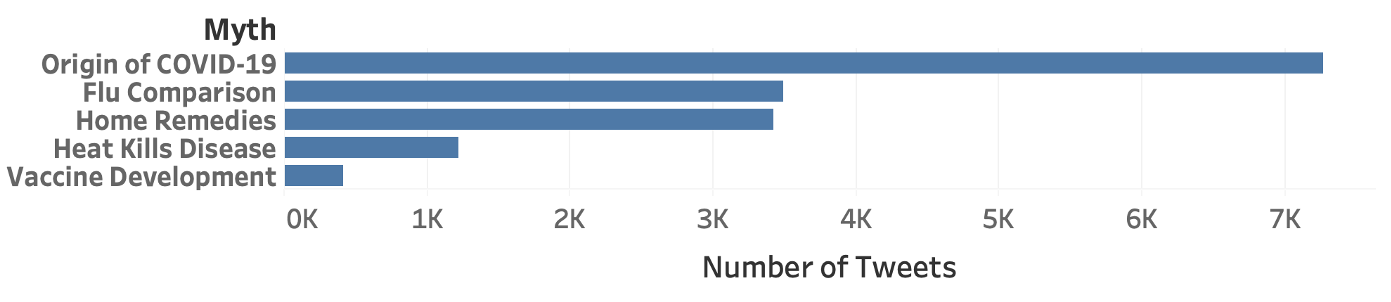}
    \caption{Distribution of by Myths}
    \label{fig:myth_proportion}
\end{figure}

Focusing on the number of tweets originally containing one or more of these five myths, we find that approximately 16,000 tweets (just under 0.6\% of the tweets) are discussing one or more of the myths we consider. Figure \ref{fig:myth_proportion} shows the distribution of the volume of each myth discussed. 

\begin{figure}
\centering
\includegraphics[width=0.7\linewidth]{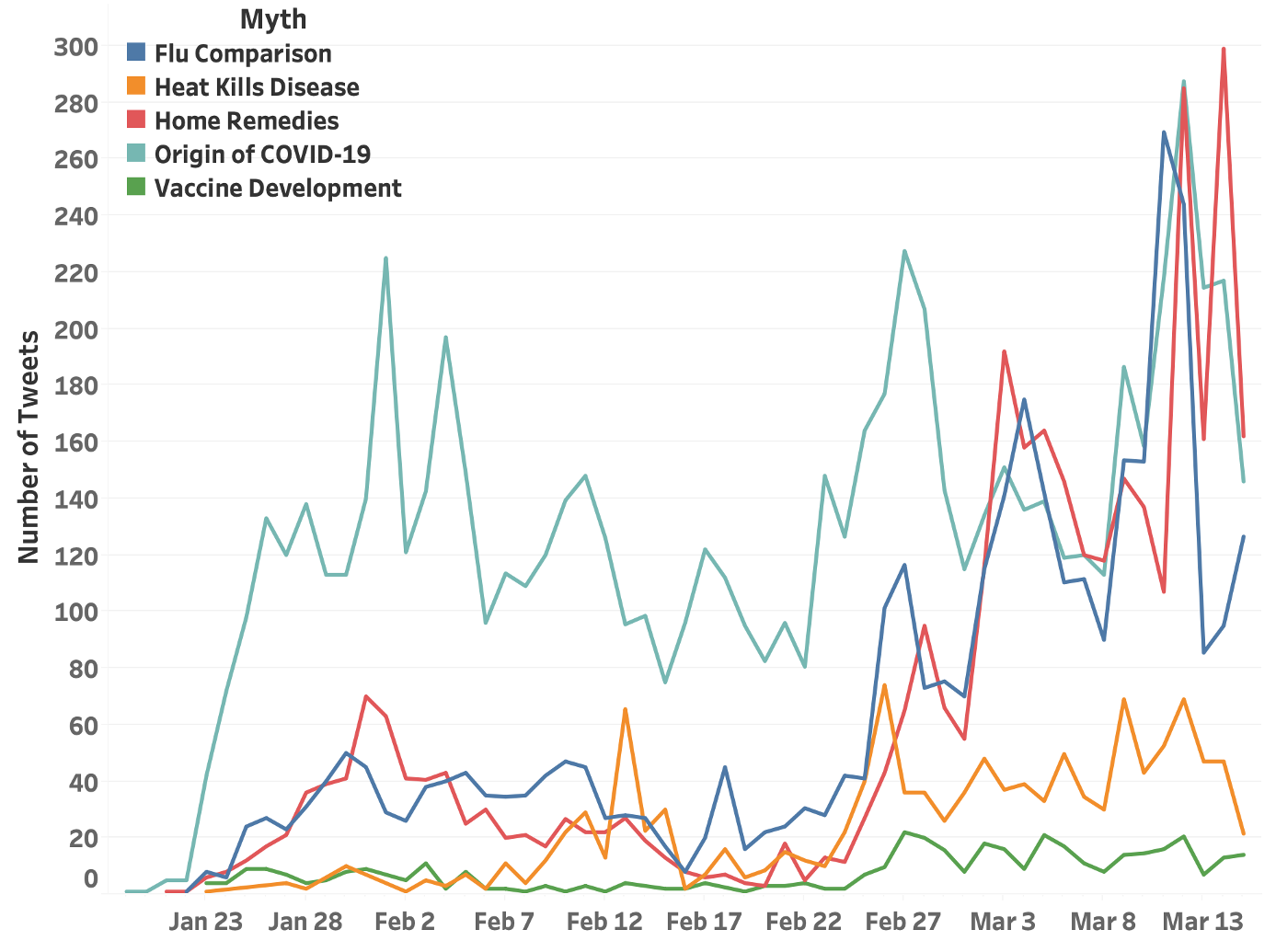}
\caption{Tweet Volume by Myths Over Time}
\label{fig:myth_trend}
\end{figure}

Tweet volume associated with the myths in our sample has increased since January, although this may reflect the growing conversation surrounding COVID-19 generally, rather than a greater proportion of attention devoted to these topics (see Figure \ref{fig:myth_trend}). However, some trends are growing more than others. Specifically, the myth regarding the origins of the virus dominated the myths present on Twitter in January and February. By the end of February, however, the flu comparison myth and the home remedy myth appear almost equally as often in our data set, although there is some suggestive evidence that the flu comparison myth may have dropped off at the end of our data collection in mid-March. While myths about heat killing COVID-19 and vaccine development also appear to rise over time in our data set, they maintain their relative position compared to the other myths and may simply reflect growing conversation surrounding the topic. It is important to note that tweets that are countering or debunking a particular myth are also likely to be part of this set of identified tweets. Therefore, the numbers we are reporting here likely capture tweets that perpetuate a myth, as well as those combating it. In future analyses, we will use stance detection methods to determine the position of the author with regards to the myth.

\section{Sharing of High vs Low Quality Information on Twitter}
Social media users commonly rely on external information to convey ideas, support claims, and serve information needs. Social media use around COVID-19 is no exception. Our analysis of tweets related to the disease show that 40.5\% of original tweet content (5.1\% of the retweet content and 9.6\% of the overall content) includes a URL. Overall, we think the very high percentage of URLs reflects the incredible need for information in this uncertain time. Uncertainty is strongly related to information seeking, and this has been shown specifically in the realm of health information seeking and sharing online
\cite{Lin.2016}. 

\subsection{Popular domains}
We begin by examining these shared links to determine the most popular domains. There are over sixty thousand unique domains that people share in their tweets. Table~\ref{table-top-10-domains} presents the top-10 domains with respect to their frequency in tweets having more than 100 user accounts tweeting the domain.

\begin{table}[h]
\centering
\caption{Top-10 domains mentioned in tweets by over 100 handles.}
\label{table-top-10-domains}
\begin{tabular}{|l|c|}
\hline
\rowcolor[HTML]{CBCEFB} 
\multicolumn{1}{|c|}{\cellcolor[HTML]{CBCEFB}\textbf{Domain}} & \textbf{Tweet Frequency} \\ \hline
youtu.be & 78,784 \\ \hline
youtube.com & 21,894 \\ \hline
instagram.com & 19,158 \\ \hline
nytimes.com & 13,678 \\ \hline
bit.ly & 12,816 \\ \hline
scmp.com & 11,897 \\ \hline
amzn.to & 11,868 \\ \hline
theguardian.com & 11,671 \\ \hline
bbc.com & 10,220 \\ \hline
gisanddata.maps.arcgis.com & 9,181 \\ \hline
\end{tabular}
\end{table}

Inspecting these top-10 domains reveals a number of important points.  First, people are linking to YouTube a lot. The two different manifestations of YouTube - “youtu.be” and “youtube.com” - are collectively linked to in almost 100,000 tweets. Top-10 sites include news media sites from the U.S. (the New York Times), news media sites from China (South China Morning Post, scmp.com), and retail sites (Amazon). This highlights the diversity of the nature and quality of information shared from external sources on Twitter about COVID-19.

If we look at domains that  have a similar volume to our top-10 domains, but are tweeted by a small number of handles, three domains fall into this category, indicating that some domains/users are attempting to appear more relevant to the conversation.
For example, thepigeonexpress.com, a dubious site that claims to be “focused on news that’s not on mainstream media.”\footnote{\url{https://thepigeonexpress.com/about/}} has almost as many shares as the New York Times, but less than 20 accounts posting articles using this domain. Given this finding, we want to better understand whether social media users are referencing reputable sources or questionable ones? 

\subsection{High health and low quality information sources}
To answer this question, we examine the URLs being shared in our data set to determine whether or not more low quality links are being shared compared to high quality ones. 

\subsubsection{Methodology}
We begin by identifying a set of high quality health sources and a set of low-quality/questionable content providers. 

\textit{High Quality Health Sources (HQHS)}: We identify the set of reputable web domains that publish health information as follows. We first identify all countries identified by the CDC as a Level 3 travel health notice country (that is, with the recommendation to “avoid all non-essential travel”). For each of these countries, we identify the webpage (domain) of each country’s equivalent to a center for disease control. Next, we augment this list by including top medical journals and hospitals, and by identifying additional US government agencies that had official COVID-19 related recommendations (for example, while not a public health organization, the EPA released information about disinfectants that were effective). After the White House announcement regarding the America’s Health Insurance Plan’s collaboration with the White House Coronavirus Task Force, the AHIP Statement page clarifying the free testing plan was also included on this list.

\textit{Low-quality Misinformation Sources (LQMS)}: Misinformation is prevalent online. While the 2016 election brought attention to this issue, it is nothing new--especially in the health domain. For instance, a 2010 study by~\cite{Scanfeld.2010} examined 1000 randomly selected tweets mentioning antibiotics and found that 700 of them contained medical misinformation or malpractice. We identify the set of low-quality/questionable sources of information using a list curated by NewsGuard~\cite{Newsguard.03232020}. NewsGuard is a journalistic organization that generally rates websites on their tendency to spread true or false information. Since the COVID-19 outbreak, they have kept a separate list of websites identified as propagating misinformation specifically related to the virus. 

\subsubsection{Findings}
Table~\ref{table-count-tweets-by-types} shows the number of tweets containing HQHS URL links and LQMS URL links. A small fraction of the urls shared come from one of the three categories listed above. The tweets containing a link to a Reputable Health Sources account for 0.51\% of tweets and 0.04\% of retweets. Low-quality/Questionable Sources account for 0.4\% of original tweets and 0.06\% of retweets. We see that both HQHS and LQMS sites are shared very infrequently, with LQMS being tweeted at a lower rate than HQHS. LQMS are being retweeted at a rate higher than HQHS. This is concerning, but still a small fraction of the overall conversation volume.

\begin{table}[h]
\centering
\caption{Count of Tweets, Retweets, and Quote Tweets by Types of URLs.}
\label{table-count-tweets-by-types}
\resizebox{\textwidth}{!}{%
\begin{tabular}{|l|c|c|c|c|c|}
\hline
\rowcolor[HTML]{CBCEFB} 
\multicolumn{1}{|c|}{\cellcolor[HTML]{CBCEFB}\textbf{Data set}} & \textbf{Tweet Count} & \textbf{\begin{tabular}[c]{@{}c@{}}Tweets\\ with URLs\end{tabular}} & \textbf{\begin{tabular}[c]{@{}c@{}}HQHS\\ URL Count\end{tabular}} & \textbf{\begin{tabular}[c]{@{}c@{}}LQMS\\ URL Count\end{tabular}} & \textbf{Both Count} \\ \hline
Original Tweets & 2,792,513 & 1,131,112 & 14,485 & 11,654 & 8 \\ \hline
Quotes & 456,878 & 16,200 & 594 & 175 & 0 \\ \hline
Retweets & 18,168,161 & 926,103 & 8,813 & 11,415 & 0 \\ \hline
Combined & 21,417,552 & 2,073,415 & 23,892 & 23,244 & 8 \\ \hline
\end{tabular}%
}
\end{table}

\subsection{News media information sharing}
Given the infrequency of links to HQHS sites and LQMS sites, we turn our attention to news sources. News media plays an important role in informing the public. This role is heightened in crisis events such as pandemics. These organizations employ fact checkers, engage with experts with relevant expertise, and are therefore sources for trustworthy information. As such, we expect them to play a significant role in the information produced and shared on social media platforms. 

\subsubsection{Methodology}
To identify reputable news sources, we adopt the definition and list of traditional news sites shared by MediaBias/FactCheck– an independent online media outlet maintained by a small team of researchers and journalists~\cite{MediaBiasFactCheck.2018}. This list has over one thousand three hundred web domains listed as reliable news sources. 

For our twitter data set, we download all of the urls that are shared between January 16, 2020 to March 15, 2020. We then download the news articles and scrape the web content to determine the sources the news articles are referencing by inspecting the links they include (e.g. a link to the CDC site in a particular New York Times article). We then aggregate this information to determine the fraction of HQHS sites and LQMS sites referenced by the news media. 

\subsubsection{Findings}
\begin{figure}
    \centering
    \includegraphics[width=0.9\linewidth]{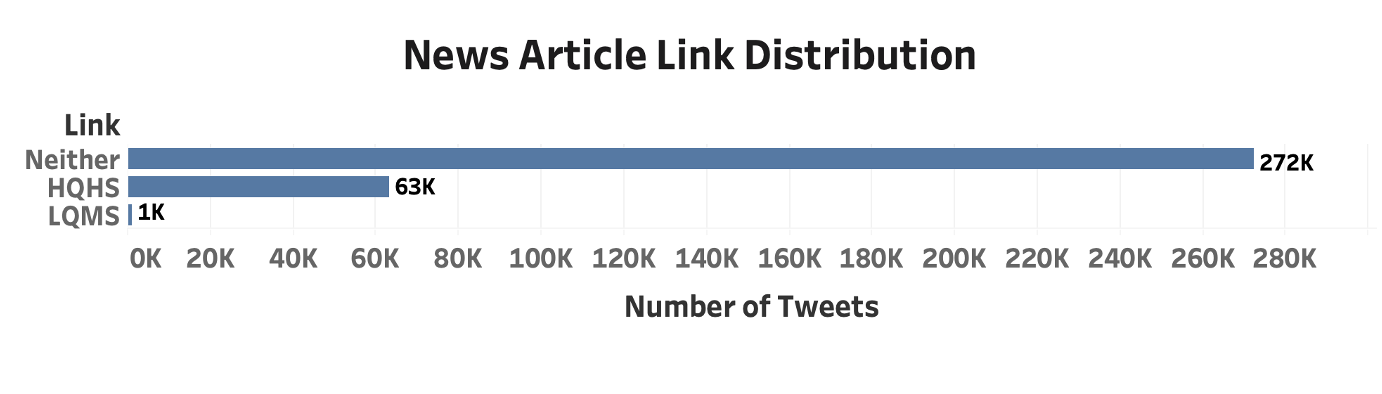}
    \caption{Distribution of Links from News Articles}
    \label{fig:news_link}
\end{figure}

Just over 351,000 of the tweets contain links to news organizations. This represents 13\% of the original tweets in our sample.\footnote{We were unable to download approximately 4\% of the HTML content for the news links.} When checking the news articles for links to high and low quality sources, we find that over 63,000 of the tweets have links to high quality sources and over 1,000 have links to low quality sources (see Figure \ref{fig:news_link}). This indicates that 18\% of the news shares are connected to HQHS sites and less than 0.3\% link to LQMS.

Focusing on more frequently shared news sources (news domains), we find that 228 new-domains were mentioned in at least 100 tweets. Of those, there are 178 news domains with at least one article that links to at least one HQHS site or LQMS site. We inspect this subset and see that 175 out of 178 are referring to HQHS site at least 80\% of the time. There are only three sites that are below that 80\% rate (newyorker.com with a 0.17 high quality to high + low quality link ratio, liveleak.com with a 0.33 ratio, and thepostmillennial.com with a 0.5 ratio). 

For the long tail of less popular domains, the results are somewhat comparable. There are 352 domains with at least one article containing at least one link to a  HQHS site or LQMS site. Out of those, only 16 have a high quality to high quality + low quality ratio lower than 0.8. Overall, these numbers are encouraging. However, it is important to note that this analysis applies only to articles that link to other sources in our data set. 

\section{Discussion and conclusions}

Our results provide important implications for understanding disease spread, information seeking behaviors during public health crises, and general communication patterns in this unprecedented combination of global pandemic and modern information environment. 

First, our results show that the COVID-19 cases are highly correlated with Twitter conversations; and further, Twitter conversations led the COVID-19 cases by 2-5 days. This is an important finding, suggesting that we can use Twitter conversations to help predict the spread and outbreak of COVID-19 when other reliable leading indicators are not available. While measurement error is of concern given this non-representative population, developing reliable ways to measure and re-weight for different biases present in these organic data sets is an important future direction. 

Second, we have learned a great deal about how much, where, and how people are communicating about this pandemic. Attention to COVID-19 continues to grow on Twitter, and likely on other platforms as well. As we know from research, people tend to care about news that affects them personally \cite{Edgerly.2018}, so it makes sense that relevant conversation would grow as the pandemic continues to affect more and more people on a personal level. Likewise, attention is focused in the countries that have been hardest hit by the disease, again suggesting that attention, discussion, and information sharing are greatest for those who are most impacted. 

But not all of this information is reliable - although people are sharing many URLs (with 40.5\% of original tweets including a URL), they are sharing less from very credible health sources like the CDC and WHO than might be expected (only 14,485 original tweets (0.4\%) did so). However, sources that we can confidently label as producers of misinformation (through domains identified as purveyors of misinformation by a journalism organization) are also not shared in great numbers (11,654 original tweets linked to such a source), even though they are retweeted more often (11,415) than credible health sources (8,813). Of course there may be - and likely are - pieces of misinformation being shared without links, or with links outside of the list of confirmed dubious URLs we used.  This preliminary analysis, though, suggests that credible information is roughly keeping pace with misinformation. 

This finding is amplified if we consider links to news sources as well. Linking to news sources was much more common than sharing either high quality health sources or known low credibility sites, with over 350,000 tweets doing so. Consistent with our expectations, these news links are much more likely to themselves link to credible sources like the WHO (63,352 news articles) than to misinformation sources (about 1,135 news articles), meaning people are likely getting reliable information from these tweets. Future work will consider all the shared content through URLs to determine the distribution of high and low quality links across all of the URL content. 

Our analysis of common myths about COVID-19 reveals a similar pattern. We started from a set of 10 different themes of myths and were able to identify five with high precision. Analysis of the tweets that pertain to these myths show that they account for a small fraction of Twitter content. While this finding is encouraging, there are many more myths that needs to be analyzed to understand the total impact of myths on the COVID-19 conversation. In general, it is crucial to continually monitor myths in the conversation, differentiate those who are propagating the myth and those who are debunking it, and build systems to combat them.

We pause to mention that many of the text analysis methods used were simple methods. Because of the volume of data and in the interest of time, we chose to use these simpler techniques in this first analysis. In future work, we will compare these simple approaches to more robust analysis techniques to improve our overall understanding of these data. 

While we have mentioned future directions throughout the article, we highlight the ones we are already making progress on. One direction of future work will focus on conducting a more refined spatio-temporal analysis of the flow of information and the transmission of COVID-19. We also intend to use language models in conjunction with machine learning models to identify myths and themes with high levels of recall and precision. Finally, as the crisis continues to unfold, we will work to share results and aggregate data sets through a web portal for those interested in advancing research related to COVID-19 and social media conversation.

\bibliographystyle{tfnlm}
\bibliography{COVID2020}

\section*{Acknowledgements}

This research was primarily supported by the Massive Data Institute at Georgetown University. It was also supported by the Computational and Spatial Analysis Core at Pennsylvania State University,  the National Science Foundation (Awards SES-1823633,  SES-1934925, SES-1934494, CNS-1453392), the Eunice Kennedy Shriver National Institute of Child Health and Human Development (Awards P2C HD041025), the USDA National Institute of Food and Agriculture and Multistate Research Project \#PEN04623 (Accession \#1013257).

\vspace{3mm}
\noindent We would also like to acknowledge the support of others on our team:\\
\hspace{3mm}- Junjun Yin at Penn State extracted geotagged tweets.\\
\hspace{3mm}- Yiqing Ren at Georgetown computed volumes.\\
\hspace{3mm}- Virinche Marwadi at Georgetown for adjusting infrastructure.\\ 
\hspace{3mm}- Robert Churchill helped with myth identification.

\newpage
\section{Appendix A}

\begin{table}[h]
\centering
\caption{Hashtags and Start Date of Collection.}
\label{table-hastags}
\begin{tabular}{|l|c|}
\hline
\rowcolor[HTML]{CBCEFB} 
\multicolumn{1}{|c|}{\cellcolor[HTML]{CBCEFB}\textbf{Hashtag}} & \textbf{\begin{tabular}[c]{@{}c@{}}Date of First\\ Collection\end{tabular}} \\ \hline
\#2019nCoV & 1/16/2020 \\ \hline
\#ChinaPneumonia & 1/16/2020 \\ \hline
\#ChinesePneumonia & 1/16/2020 \\ \hline
\#Corona & 1/16/2020 \\ \hline
\#SARI & 1/16/2020 \\ \hline
\#WuhanCoronavirus & 1/16/2020 \\ \hline
\#WuhanPneumonia & 1/16/2020 \\ \hline
\#CoronavirusOutbreak & 1/20/2020 \\ \hline
\#VirusChina & 1/21/2020 \\ \hline
\#Coronavirus* & 1/23/2020 \\ \hline
\#Wuhan & 1/23/2020 \\ \hline
\#CoronaOutbreak & 3/02/2020 \\ \hline
\#COVID & 3/02/2020 \\ \hline
\#CoronavirusUpdate & 3/11/2020 \\ \hline
\#COVID\_19 & 3/11/2020 \\ \hline
\#COVID19** & 3/11/2020 \\ \hline
\end{tabular}
\end{table}


\end{document}